\documentclass{aa}

\usepackage{graphicx}
\usepackage{txfonts}
\usepackage{threeparttable}  
\usepackage{booktabs}
\usepackage{natbib}
\usepackage{makecell}
\usepackage[colorlinks=true, linkcolor=blue, citecolor=blue, filecolor=blue, urlcolor=blue]{hyperref}

\begin{document} 

\title{Direct VLBI evidence for a buried AGN in the triple-merger LIRG UGC~2369S}

\authorrunning{W. Xu et al.}
\titlerunning{Direct VLBI evidence for a buried AGN in the triple-merger LIRG UGC~2369S}

\author{
    Wancheng Xu
    \inst{1,2,3} \email{xuwancheng@xao.ac.cn}
\and
    S\'andor Frey
    \inst{1,4} \email{frey.sandor@csfk.org} 
\and
    Lang Cui
    \inst{3,2} \corrauth{cuilang@xao.ac.cn}
\and
    Krisztina Éva Gabányi
    \inst{4,5,1} \email{k.gabanyi@astro.elte.hu}
}

\institute{
    Konkoly Observatory, HUN-REN Research Centre for Astronomy and Earth Sciences, MTA Centre of Excellence, Konkoly Thege Mikl\'os \'ut 15-17, H-1121 Budapest, Hungary    
\and
    School of Astronomy and Space Science, University of Chinese Academy of Sciences, No. 1 Yanqihu East Road, Beijing 101408, China
\and
    Xinjiang Astronomical Observatory, Chinese Academy of Sciences, 150 Science 1-Street, Urumqi 830011, China
\and
    Department of Astronomy, Institute of Physics and Astronomy, ELTE E\"otv\"os Lor\'and University, P\'azm\'any P\'eter s\'et\'any 1/A, H-1117 Budapest, Hungary
 \and
    HUN-REN--ELTE Extragalactic Astrophysics Research Group, ELTE E\"otv\"os Lor\'and University, P\'azm\'any P\'eter s\'et\'any 1/A, H-1117 Budapest, Hungary
}

\date{Received 04 May 2026 / Accepted 26 June 2026}

\abstract
{UGC~2369S is a luminous infrared galaxy (LIRG) undergoing a late-stage merger in a triple system, where the heavily obscured northern core is suspected to host an active galactic nucleus (AGN). However, severe dust and gas obscuration makes definitive confirmation challenging.}
{We aim to provide direct observational evidence for the buried AGN through high-resolution radio imaging, while investigating the AGN accretion and feedback properties within this merger-driven gas-rich environment.}
{We analyzed archival European VLBI Network (1.6~GHz) and Very Long Baseline Array (1.7 and 5~GHz) data of UGC~2369S. Through high-resolution imaging and visibility-domain Gaussian modeling, we characterized the morphology and intensity of its milliarcsecond-scale radio emission.}
{A compact radio component is detected at the northern core, exhibiting high brightness temperature ($T_{\rm b}>10^7~\rm K$) and flat radio spectrum ($\alpha \approx-0.45$), which confirms the presence of an obscured AGN. The sub-Eddington accretion rate ($\lambda_{\rm Edd} \approx 2.7 \times 10^{-4}$) indicates that it falls within the radiatively inefficient accretion flow (RIAF) state.}
{We provide direct imaging evidence for an AGN in the northern core of UGC~2369S, revealing a deeply buried, jet-emitting low-luminosity AGN (LLAGN) enshrouded by a Compton-thick gas cocoon. This demonstrates that VLBI is a uniquely effective tool for disentangling nuclear accretion and feedback processes within the heavily obscured environments of multiple-merger systems.}

\maketitle
\nolinenumbers

\section{Introduction}\label{section_1}

Galaxy mergers play a fundamental role in the co-evolution of supermassive black holes (SMBHs) and their host galaxies \citep{1988ApJ...325...74S,2008ApJS..175..356H}. During late-stage mergers, strong gravitational torques effectively remove angular momentum from the gas, driving massive inflows into the central regions \citep{1991ApJ...370L..65B,1996ApJ...471..115B}. These inflows not only trigger intense starbursts \citep{1996ApJ...471..115B} but also fuel the central engines \citep{2005Natur.433..604D,2012ApJ...758L..39T}, frequently forming dual or multiple active galactic nuclei (AGNs) when two or more SMBHs are simultaneously active \citep[e.g.,][]{2012ApJ...746L..22K}. Among them, triple-merger systems represent an exceptionally rare and extreme regime, providing unique insights into AGN accretion and feedback in complex environments \citep{2019NewAR..8601525D,2019ApJ...883..167P}. However, these merger-driven inflows create remarkably dense environments, heavily obscuring the accreting SMBHs \citep{2017MNRAS.468.1273R,2018Natur.563..214K}. In the case of jetted AGNs, very long baseline interferometry (VLBI) provides a highly effective approach to probe the exact interplay among accretion triggering, nuclear obscuration, and jet feedback in multi-SMBH systems \citep[e.g.,][]{2014Natur.511...57D,2018RaSc...53.1211A,2019A&A...630L...5G}.

\begin{table*}[htbp]
\centering
\small
\begin{threeparttable}
\caption{VLBI observing information for UGC~2369S.} 
\label{table_1}
\begin{tabular}{ccccccc}
\toprule\midrule
Array & Project ID & Date & Frequency & Phase calibrator & Fringe finder & Antenna \\
 (1)  &    (2)     & (3)  &    (4)    &       (5)        &      (6)      &   (7)   \\
\midrule
EVN  & EC020B & 2003 May 27 & 1.6~GHz & J0256+1334 & DA193      & ON, WB, MC, EF, NT, JB, TR, CM         \\ 
VLBA & BL029  & 1996 Jan 17 & 1.7~GHz & J0238+1636   & J0238+1636   & BR, FD, HN, KP, LA, MK, NL, OV, PT, SC \\
VLBA & BM190A & 2003 Jul 17 & 5.0~GHz & J0256+1334 & J0050$-$0929 & BR, FD, HN, KP, LA, MK, NL, OV, PT, SC \\  
\bottomrule
\end{tabular}
\begin{tablenotes}
\item {\bf Note.} Antenna name: ON -- Onsala (Sweden), WB -- Westerbork (phased array; The Netherlands), MC -- Medicina (Italy), EF -- Effelsberg (Germany), NT -- Noto (Italy), JB -- Jodrell Bank Lovell (Great Britain), TR -- Toru\'n (Poland), CM -- Cambridge (Great Britain), BR -- Brewster, FD -- Fort Davis, HN -- Hancock, KP -- Kitt Peak, LA -- Los Alamos, MK -- Mauna Kea, NL -- North Liberty, OV -- Owens Valley, PT -- Pie Town, SC -- St. Croix (all USA).
\end{tablenotes}
\end{threeparttable}
\end{table*}

UGC~2369S, the southern irregular galaxy in the UGC~2369 interacting pair, is a late-stage luminous infrared galaxy (LIRG) merger system comprising three distinct components separated by $\sim 3$~kpc: the northern (N), southeast (SE), and southwest (SW) cores (see Figure~\ref{figure_A1}; \citealt{2010AAS...21560203R,2026ApJ..1002..130D}). Recent comprehensive multi-wavelength observations characterized this complex triple-merger environment, indicating that a buried AGN might reside within the northern core \citep{2026ApJ..1002..130D}. However, definitive observational confirmation of the central engine is still lacking, as the overall infrared emission can also be fully explained by starburst activity \citep{2008A&A...484..631V}. To unambiguously identify the central engine, high-resolution radio observations are crucial. While early VLBI snapshot detection experiments inferred an AGN-like core with a high brightness temperature ($>10^7$~K) in UGC~2369S \citep{1993ApJ...405L...9L} and indirectly deduced its presence by ruling out extreme radio supernova models \citep{1998ApJ...492..137S}, direct imaging evidence, unambiguous spectral confirmation, and the precise localization of the radio core remained absent.

To address this, we analyze archival VLBI data in this paper to reveal the parsec-scale radio structure of UGC~2369S, providing direct evidence for a buried AGN within the dusty starburst environment. Section~\ref{section_2} details the VLBI data reduction, followed by the imaging results in Section~\ref{section_3}. In Section~\ref{section_4}, we present the multi-wavelength analysis and discussions. A summary of our conclusions is provided in Section~\ref{section_5}. Throughout this work, a flat $\Lambda$CDM cosmological model with $\Omega_{\rm m}=0.31$, $\Omega_{\rm \Lambda}=0.69$, and $H_0=67.7$~km~s$^{-1}$~Mpc$^{-1}$ is adopted \citep{2020A&A...641A...6P}. The redshift of the source is $z=0.0318$ \citep{2026ApJ..1002..130D}, corresponding to a luminosity distance of $D_{\rm L}=144.2$~Mpc, yielding a physical scale of $\approx0.66~{\rm pc}$ per milliarcsecond (mas). To facilitate comparison with the literature, we report radio luminosities in SI units (W or $\rm W~Hz^{-1}$) and luminosities in X-ray and optical bands in CGS units ($\rm erg~s^{-1}$).

\section{Data reduction}\label{section_2}

In this study, we retrieved the publicly available archival VLBI data for UGC~2369S from the European VLBI Network (EVN; comprising eight European antennas) and the Very Long Baseline Array (VLBA). The dataset comprises three projects: EC020B (PI: J.E. Conway, mentioned in \citealt{2004ASPC..320..226P}), BL029 (PI: C.J. Lonsdale), and BM190A (PI: R. Maiolino). The total on-source time for the target UGC~2369S, observed in phase-referencing mode, was approximately 100 minutes in each project. Due to technical malfunctions and tape recording errors at certain stations, a portion of the data from project EC020B was irreversibly lost, resulting in a slight degradation of the overall array sensitivity and $(u,v)$-coverage. Table~\ref{table_1} summarizes the observational details, while Figure~\ref{figure_A1} displays the corresponding fields of view (FoVs). Although these observations were conducted more than 20 years ago with the primary goal of studying elusive AGNs in starburst galaxies, to our knowledge, the data have not been published until now. 

Data calibration was performed using the US National Radio Astronomy Observatory (NRAO) Astronomical Image Processing System (AIPS; \citealt{greisen2003aips}). For the VLBA data, the calibration generally followed the standard procedures described in the AIPS Cookbook\footnote{\url{http://www.aips.nrao.edu/cook.html}}; the EVN data were calibrated following the EVN Data Reduction Guide\footnote{\url{https://www.evlbi.org/evn-data-reduction-guide}}. A priori amplitude calibration was applied using the antenna system temperatures (TY) and gain curves (GC) provided by the VLBI stations. Following instrumental and bandpass corrections, global fringe-fitting solutions derived from the phase calibrators were interpolated and applied to the target source. For the 1996 observation (project BL029), ionospheric correction was not performed because the ionospheric calibration file was unavailable for this early epoch. The Los Alamos (LA) and Effelsberg (EF) stations were chosen as the reference antennas for the VLBA and EVN data calibration, respectively.

The calibrated visibility data were loaded into the DIFMAP software package \citep{1997ASPC..125...77S} for imaging and brightness distribution modeling. An initial inspection of the dirty maps revealed a clear signal ($>8\sigma$) near the northern core position in all three observations. The positional stability of this signal under different weighting schemes (i.e., uniform and natural weighting) further demonstrates the robustness of the detection. Phase and amplitude self-calibration was not performed given the faintness of the detected core ($S_{\nu}<5~{\rm mJy}$), resulting in conservative flux density estimates due to coherence loss \citep[e.g.,][]{2006A&A...445..413M,2010A&A...515A..53M}. To quantitatively describe the source structure, we fitted a single circular Gaussian model to the visibility data of each observation using the \texttt{modelfit} task, deriving the component size, position, and flux density.

\section{Results}\label{section_3}

\begin{figure*}[htbp]
\centering
\includegraphics[width=0.31\textwidth]{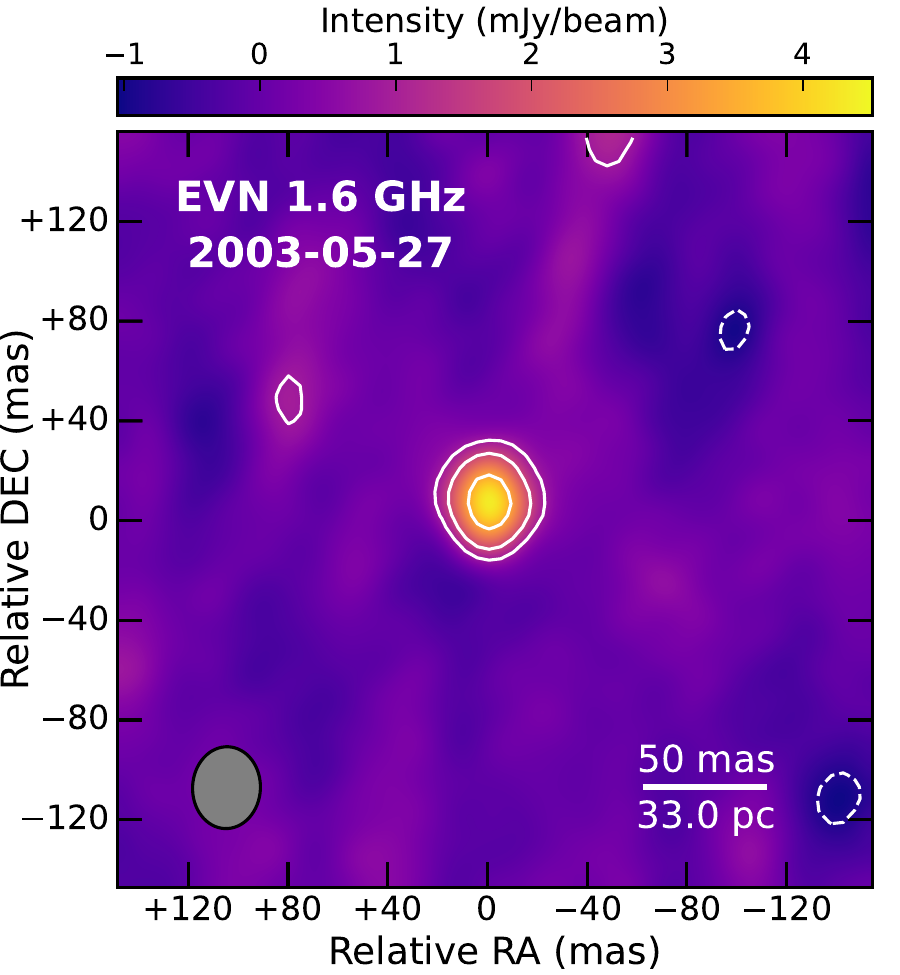}
\includegraphics[width=0.31\textwidth]{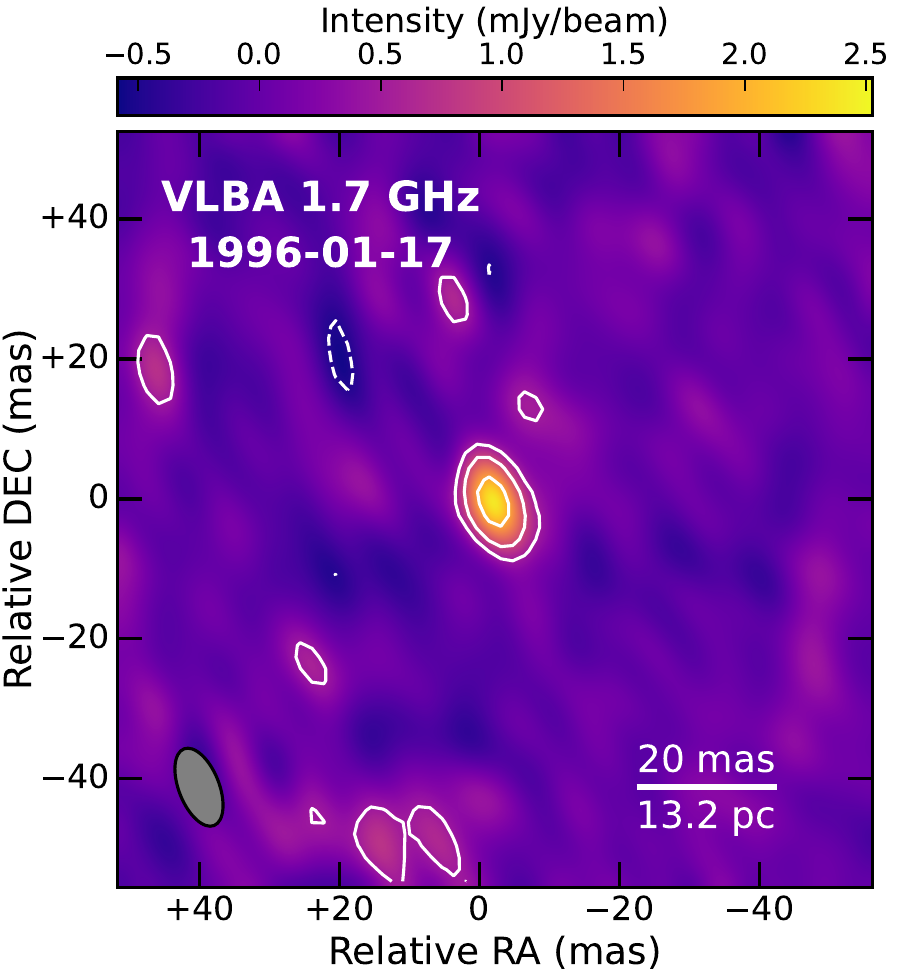}
\includegraphics[width=0.31\textwidth]{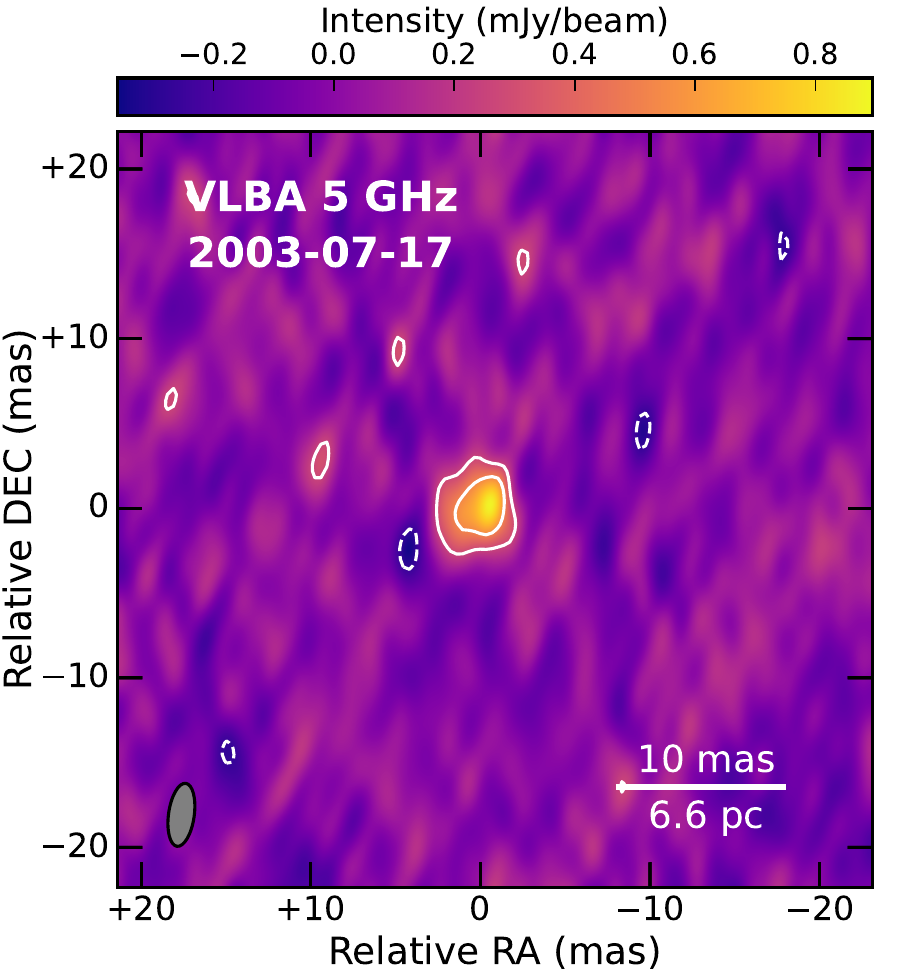}
\caption{Naturally weighted VLBI images of UGC~2369S. The central coordinates of all panels are aligned to the core position derived from the VLBA 5~GHz image. The synthesized restoring beams are shown as gray ellipses in the bottom-left corner of each panel. Scale bars are shown in the bottom right corner of each panel. The contours are drawn at $5\sigma \times [-1, 1, 2, 4, 8]$ for the EVN 1.6~GHz image, and at $3\sigma \times [-1, 1, 2, 4, 8]$ for the VLBA 1.7 and 5~GHz images. Negative contours are represented by dashed lines. The root-mean-square (rms) noise levels ($\sigma$) determined from the residual maps are 0.172, 0.162, and 0.088~mJy~beam$^{-1}$ for the EVN 1.6~GHz, VLBA 1.7~GHz, and VLBA 5~GHz images, respectively.}
\label{figure_1}
\end{figure*}

Figure~\ref{figure_1} shows the restored VLBI images after Gaussian modeling. A bright radio component corresponding to the northern core was clearly detected ($>18\sigma$) in all three observations. We applied an astrometric correction to the target coordinates based on updated phase calibrator coordinates from the Astrogeo RFC\_2026a release (\citealt{petrov2025}; see Appendix~\ref{appendix_B}). The corrected VLBI coordinates, along with the Gaia DR3 counterpart \citep{Gaia2016,Gaia2023}, are listed in Table~\ref{table_2}. Among them, the VLBA 5~GHz coordinates represent the most precise position of the AGN jet base. The radio positional uncertainties account for statistical errors from thermal noise, intrinsic uncertainties of the phase calibrator, and an additional systematic error of $\sim$1~mas arising from unmodeled atmospheric phase residuals scaled by the target--calibrator separation \citep{pradel2006astrometric}. Notably, the spatial offsets among the different observations significantly exceed the positional uncertainties, likely arising from resolution mismatches, intrinsic source structure variations, and Gaia photocenter jitter within the dust-obscured merger environment.

\begin{table}[htbp]
\centering
\tiny
\begin{threeparttable}
\caption{Coordinates of the northern core in UGC~2369S.} 
\label{table_2}
\begin{tabular}{cccc}
\toprule\midrule
 Observation &   R.A.  &  Decl.  & Pos. error \\
             & (J2000) & (J2000) &   (mas)    \\
\midrule
EVN  1.6~GHz & $02^{\rm h}54^{\rm m}01\fs819001$ & +14\degr58\arcmin14\farcs92208 & $\pm 1.53$ \\
VLBA 1.7~GHz & $02^{\rm h}54^{\rm m}01\fs818946$ & +14\degr58\arcmin14\farcs91188 & $\pm 1.25$ \\
VLBA 5.0~GHz & $02^{\rm h}54^{\rm m}01\fs819049$ & +14\degr58\arcmin14\farcs91551 & $\pm 1.23$ \\  
Gaia DR3     & $02^{\rm h}54^{\rm m}01\fs819383$ & +14\degr58\arcmin14\farcs91108 & $\pm 0.84$ \\
\bottomrule
\end{tabular}
\begin{tablenotes}
\item {\bf Note.} For the Gaia position, the astrometric excess noise (AEN) is adopted as the positional uncertainty. 
\end{tablenotes}
\end{threeparttable}
\end{table}

The VLBI imaging and model parameters of the northern core in UGC~2369S are listed in Table~\ref{table_3}, with detailed calculation processes provided in Appendix~\ref{appendix_C}. The circular Gaussian modeling reveals a compact structure across all observations, remaining unresolved at 1.6 and 1.7~GHz, but partially resolved in the VLBA 5~GHz image. Source flux densities were derived from Gaussian modeling without self-calibration to prevent the creation of spurious signals. Considering potential coherence loss from residual atmospheric phase errors \citep[e.g.,][]{2006A&A...445..413M,2010A&A...515A..53M}, the values reported in Table~\ref{table_3} represent conservative estimates of the intrinsic core emission. The uncertainties were calculated by adding a $10\%$ systematic calibration error in quadrature to the $1\sigma$ image rms noise, i.e., $\sigma_{S}=\sqrt{0.01S_{\nu}^2+{\rm rms}^2}$. 

For the $L$--$C$ band two-point power-law spectral index $\alpha$ ($S_{\nu} \propto \nu^{+\alpha}$), the non-simultaneous VLBA 1.7 and 5~GHz data yield $\alpha_{1.7}^{5.0} = -0.45 \pm 0.14$. Alternatively, incorporating the quasi-simultaneous ($\Delta t < {\rm 60~days}$) EVN 1.6~GHz data results in a steeper spectrum ($\alpha_{1.6}^{5.0} = -0.73 \pm 0.13$). This steepening can be attributed to the larger restoring beam of the EVN (eight intra-European antennas with a maximum baseline length of $\sim2300$~km), which captures extended emission resolved out by the longer VLBA baselines (maximum of $\sim 8500$~km). Indeed, applying a comparable long-baseline cutoff ($\sim 12.2\ \mathrm{M}\lambda$) to the VLBA 1.7~GHz data recovers a higher flux density ($\sim 4.05$~mJy) that is in better agreement with the EVN measurement. Furthermore, our $L$-band model flux densities align well with the correlated flux density of $\sim 5$~mJy reported in the 1991 global-VLBI snapshot \citep{1993ApJ...405L...9L,1998ApJ...492..137S}, indicating no direct evidence for significant long-term AGN variability. Given the better-matched beam sizes of the two VLBA epochs, we adopt the flatter spectrum ($\alpha \approx -0.45$) as a more accurate representation of the compact core. In addition, the core exhibits high brightness temperatures ($T_{\rm b}>10^7$~K) and radio monochromatic luminosities ($L_\nu>10^{21}~{\rm W~Hz}^{-1}$) characteristic of faint Seyfert cores or low-luminosity AGNs \citep[LLAGNs,][]{2005A&A...435..521N,2008ARA&A..46..475H}. Together, these properties provide unambiguous evidence for a compact, non-thermal jet base. An alternative origin, such as a nuclear starburst \citep[e.g.,][]{1992ARA&A..30..575C, 2004A&A...417..925M} or clustered radio supernovae, can be ruled out given the unphysical source concentration required \citep{1998ApJ...492..137S}, the decade-long stability of the high brightness temperature \citep{2010A&A...519L...5P}, and the insufficient local star formation rate ($\sim 14.7~M_{\odot}~{\rm yr^{-1}}$; \citealt{2026ApJ..1002..130D}) to power the compact radio emission.

\begin{table*}[htbp]
\centering
\small
\begin{threeparttable}
\caption{VLBI imaging and model parameters of the northern core in UGC~2369S.} 
\label{table_3}
\begin{tabular}{ccccccccc}
\toprule\midrule
 Observation & Beam size & $d_{\rm lim}$ & $\theta$ & $S_{\rm peak}$ & $\sigma$ & $S_{\nu}$ & $T_{\rm b}$ & $L_{\nu}$ \\
    & (mas $\times$ mas) & (mas) & (mas) & (mJy beam$^{-1}$) & (mJy beam$^{-1}$) & (mJy) & ($10^7~\rm K$) & ($10^{21}~\rm W~Hz^{-1}$)\\
(1) & (2) & (3) & (4) & (5) & (6) & (7) & (8) & (9) \\
\midrule
EVN  1.6~GHz & $27.8\times26.3$ & 12.0 & $<$12.0 & 4.51 & 0.172 & $4.80\pm0.51$ & $>$$1.64$ & $11.9\pm1.3$ \\
VLBA 1.7~GHz & $11.8\times5.77$ & 4.21 & $<$4.21 & 2.67 & 0.162 & $3.42\pm0.38$ & $>$$8.40$ & $8.51\pm0.95$ \\
VLBA 5.0~GHz & $3.75\times1.52$ & 1.15 & $2.74\pm0.15$ & 0.89 & 0.088 & $2.10\pm0.23$ & $\phantom{>}1.41\pm0.22$ & $5.22\pm0.57$\\ 
\bottomrule
\end{tabular}
\begin{tablenotes}
\item {\bf Note.} Column~(1): observing array and frequency, Column~(2): restoring beam size FWHM (major axis $\times$ minor axis), Column~(3): theoretical resolution limit calculated by Equation~\ref{eq1}, Column~(4): fitted circular Gaussian component size (FWHM), with upper limits given for observations where the core remains unresolved, Column~(5): peak intensity of the restored image, Column~(6): rms noise of the residual image, Column~(7): total flux density of the fitted component, Column~(8): derived brightness temperature, with lower limits given for unresolved cores, Column~(9): intrinsic monochromatic radio luminosity at the observing frequency.
\end{tablenotes}
\end{threeparttable}
\end{table*}

\section{Discussion}\label{section_4}

Combining radio, optical, and X-ray diagnostics, we performed a multi-wavelength analysis to characterize the accretion and feedback properties of the northern core (see Appendix~\ref{appendix_D}). By applying the black hole fundamental plane (FP) relation for sub-Eddington objects \citep{2012MNRAS.419..267P} to the VLBA 5~GHz luminosity of $L_{\rm R} =\nu L_{\nu}= (2.61 \pm 0.29) \times 10^{38}~{\rm erg~s^{-1}}$ and black hole mass of $M_{\rm BH} = (4.73\pm0.24) \times 10^8 M_{\odot}$ \citep{2026ApJ..1002..130D}, we predict an intrinsic X-ray luminosity of $\log (L_{\rm X,FP}/{\rm erg~s^{-1}}) = 42.00 \pm 0.41$. The Chandra-observed X-ray luminosity of $\log (L_{\rm 2-8~keV}/{\rm erg~s^{-1}})=40.12\pm0.08$ \citep{2026ApJ..1002..130D} is lower than the VLBI-predicted value by a factor of $\sim$76, implying either an intrinsically fuel-starved engine or severe gas obscuration \citep[e.g.,][]{2013ApJ...762..110L,2024ApJ...969...36X}. Based on standard X-ray photoelectric absorption modeling (see Appendix~\ref{appendix_D1}), we estimate an extreme hydrogen column density of $N_{\rm H} \gtrsim 2.6 \times 10^{24}~{\rm cm}^{-2}$. Given that simple absorption models typically underestimate the column density under such extreme obscuration ($N_{\rm H} \gtrsim 10^{24}~{\rm cm}^{-2}$; \citealt{2009MNRAS.397.1549M}), our result is consistent with the column density independently derived from near-infrared integral field spectroscopy of the nuclear molecular gas ($N_{\rm H} \gtrsim 10^{25}~{\rm cm}^{-2}$; \citealt{2026ApJ..1002..130D}). This deeply buried nature is further corroborated by the thickness parameter ($T=L_{\rm X}/L_{\rm [O\,III]}^{\rm int}$; \citealt{1999ApJS..121..473B}). Using the observed X-ray and extinction-corrected [O~III] luminosities, the exceptionally low ratio of $T \approx 0.09$ firmly places the northern core in the Compton-thick regime ($T<1$). Ultimately, the agreement between the [O~III]-derived and FP-predicted intrinsic X-ray luminosities (see Figure~\ref{figure_D1}) confirms a powerful central engine hidden behind a dense cocoon, effectively ruling out the fuel-starved scenario.

The accretion efficiency of the northern core in UGC~2369S, characterized by the Eddington ratio, is estimated to be $\lambda_{\rm Edd} = \left(2.68^{+4.22}_{-1.64}\right) \times 10^{-4}$ (see Appendix~\ref{appendix_D2}). In the local Universe, massive black holes in isolated early-type galaxies are generally quiescent or present as faint LLAGNs ($\lambda_{\rm Edd}\lesssim 10^{-5}$, $L_{\rm 5~GHz}\sim10^{19}$--$10^{20}~{\rm W~Hz}^{-1}$; \citealt{2008ARA&A..46..475H}). Although the derived Eddington ratio for the northern core of UGC~2369S is mildly elevated compared to this quiescent baseline, its accretion state remains well within the radiatively inefficient accretion flow (RIAF) regime \citep{2008ARA&A..46..475H}. This indicates that while the central engine sustains a powerful compact jet ($L_{\rm VLBI} \sim 10^{21}$--$10^{22}~{\rm W~Hz}^{-1}$) within this advanced merger, it has not transitioned into a rapidly growing, quasar-like phase. 

For UGC~2369S, the VLA A-configuration 1.49~GHz flux density is $41.9\pm0.2$~mJy within a $1\farcs5$ aperture \citep{1990ApJS...73..359C,2015A&A...574A...4V}, which corresponds to a physical scale of $\sim 1~{\rm kpc}$, yielding a monochromatic luminosity of $L_{\rm 1.4\,GHz}\approx1.04\times10^{23}~\rm W~Hz^{-1}$ and a jet mechanical power of $L_{\rm mech}=10^{43}\times\left(\frac{L_{\rm 1.4\,GHz}}{10^{24}~{\rm W~Hz^{-1}}}\right)^{0.7}\approx2.07\times10^{42}~\rm erg~s^{-1}$ \citep[e.g.,][]{Cavagnolo2010}. Considering the bolometric luminosity of $L_{\rm bol}\approx1.60\times10^{43}~{\rm erg~s^{-1}}$, we obtain a radiation-to-mechanical power ratio of $L_{\rm bol}/L_{\rm mech} \approx 7.6$, indicating that the compact jet in the northern core of UGC~2369S carries a substantial kinetic energy fraction of $\sim 11.6\%$ (even after excluding the star formation contribution, see Appendix~\ref{appendix_D4}, it remains as high as $\sim 10\%$), which is over an order of magnitude higher than the theoretical threshold ($\sim 0.5\%~L_{\rm bol}$) required to drive effective mechanical feedback \citep{2010MNRAS.401....7H}. In this scenario, the RIAF state of the central engine naturally maintains a geometrically thick inner accretion flow, sustaining the powerful, non-thermal radio jet detected by VLBI.

On a larger scale extending to a radius of $1\farcs8$ ($\sim 1.2~{\rm kpc}$), \citet{2026ApJ..1002..130D} reported massive gas outflows with a total kinetic power of $\dot{E}_{\rm kin} \sim 3 \times 10^{42}~\rm erg~s^{-1}$, which falls well within the maximum starburst energy injection rate ($\sim 1.0 \times 10^{43}~\rm erg~s^{-1}$) and is thus consistent with a starburst-driven origin. The AGN jet mechanical power we derive ($L_{\rm mech} \approx 2.07 \times 10^{42}~\rm erg~s^{-1}$) is roughly a factor of five lower than this starburst budget, indicating that the jet alone is insufficient to power the outflow. AGN feedback can also proceed radiatively \citep[e.g.,][]{2017A&A...601A.143F}, and the AGN bolometric luminosity ($L_{\rm bol} \approx 1.60 \times 10^{43}~\rm erg~s^{-1}$) is comparable to the starburst energy injection rate. However, driving the outflow by radiation alone would require a coupling efficiency of $\dot{E}_{\rm kin}/L_{\rm bol} \approx 0.19$, above the $0.1\%$--$10\%$ range found for AGN-driven winds \citep{2017A&A...601A.143F}. Therefore, the nuclear starburst most likely dominates the energy budget of the large-scale outflows, while a contribution from the AGN cannot be fully excluded.

The analysis of radio emission origins (see Appendix \ref{appendix_D4}) indicates that the VLBI-detected emission is jet-dominated rather than coronal. Of the total VLA 1.49~GHz emission \citep{1990ApJS...73..359C,2015A&A...574A...4V}, the compact emission detected by EVN 1.6~GHz observations accounts for $\sim 11\%$, while star formation contributes an estimated $\sim 22\%$, and the residual $\sim 67\%$ is associated with extended jets or outflows resolved out by VLBI.

For the southeast (SE) and southwest (SW) components of the triple system UGC~2369S, no significant radio emission ($>6\sigma$) was detected within a $1\arcsec$ radius of their Gaia positions. Therefore, we only report $6\sigma$ upper limits for their VLBA 5~GHz flux densities. The radio luminosity upper limits and the corresponding maximum X-ray luminosities permitted by the FP relation were also estimated (see Table~\ref{table_D1}). Based on the extinction-corrected [O~III] luminosities and Equation~\ref{eq6}, the intrinsic X-ray luminosities were estimated to be $\log (L_{\rm X,[O\,III]}/{\rm erg~s^{-1})} \approx 42.24$ and 41.79 for the SE and SW cores, respectively. For the SE core, $L_{\rm X, [O\,III]} > L_{\rm X, FP}$, which implies that if it hosted a typical AGN following standard scaling relations, the expected radio emission would well exceed our VLBA 5~GHz detection threshold. The non-detections in both VLBI and Chandra observations strongly suggest that the SE core is inherently radio-deficient, where jet activity is likely suppressed by the central dense-gas environment typical of late-stage mergers. In contrast, for the SW core, assuming the nominal mass estimate (see Table~\ref{table_D1}), the intrinsic X-ray luminosity $L_{\rm X, [O\,III]}$ falls below its corresponding FP upper limit $L_{\rm X, FP}$, suggesting that the VLBI non-detection is likely due to insufficient sensitivity rather than an intrinsic deficit; thus, we cannot rule out the presence of a radio-quiet AGN. Another possibility is that the radio emission in both nuclei originates from non-AGN processes (e.g., star formation), which would be completely resolved out by the long baselines, leading to the VLBI non-detections.

\section{Conclusion and summary}\label{section_5}

We have presented high-resolution imaging and analysis of archival VLBI data to reveal the compact radio emission within the triple-merger system UGC~2369S, and investigated the accretion and feedback properties of the buried AGN in a highly gas-rich environment. The main results and conclusions of this paper are summarized below.

\begin{enumerate}
\item[1.] Our imaging of the archival EVN and VLBA data reveals a high-brightness-temperature ($T_{\rm b}>10^7~\rm K$) and flat-spectrum ($\alpha \approx-0.45$) compact radio component corresponding to the northern core, providing direct evidence for the presence of a buried AGN in UGC~2369S. We also report its precise astrometric position derived from phase-referencing.

\item[2.] The buried AGN exhibits a RIAF state with an Eddington ratio of $\lambda_{\rm Edd} \approx 2.7 \times 10^{-4}$. Although this value is mildly elevated compared to typical LLAGNs, it remains well within the sub-Eddington regime, naturally sustaining the powerful compact jet within the central sub-kpc region.

\item[3.] The observed hard X-ray emission exhibits a significant deficit ($\sim 2$~dex) relative to the [O~III]-derived intrinsic luminosity, indicating intense gas absorption. The corrected X-ray luminosity agrees well with the FP prediction, confirming the presence of a Compton-thick gas cocoon surrounding the deeply buried AGN.

\item[4.] The other two cores in UGC~2369S, SE and SW, remain undetected by VLBI, implying they are either intrinsically radio-silent, powered by non-AGN processes such as star formation, or fall below the current sensitivity limit.
\end{enumerate}

\section*{Data availability}
The VLBI FITS images shown in Figure~\ref{figure_1} are available in electronic form at the CDS via anonymous ftp to \href{https://cdsarc.cds.unistra.fr/}{cdsarc.cds.unistra.fr} (\texttt{130.79.128.5}) or via \url{https://cdsarc.cds.unistra.fr/viz-bin/cat/J/A+A/}.

\begin{acknowledgements}
We thank the anonymous referee for comments and suggestions that helped improve the paper.
This work was supported by the Tianshan Talent Training Program (grant No. 2023TSYCCX0099) and the China Scholarship Council (grant No. 202504910180). 
This research was also supported by HUN-REN Hungarian Research Network.
The European VLBI Network is a joint facility of independent European, African, Asian, and North American radio astronomy institutes. Scientific results from data presented in this publication are derived from the EVN project EC020B (\url{https://doi.org/10.48717/wxd1-ar28}). The Very Long Baseline Array is an instrument of the National Radio Astronomy Observatory, which is a facility of the National Science Foundation operated under cooperative agreement by Associated Universities, Inc. This work used VLBA data from projects BL029 and BM190A. This work has made use of data from the European Space Agency (ESA) mission Gaia (\url{https://www.cosmos.esa.int/gaia}), processed by the Gaia Data Processing and Analysis Consortium (DPAC, \url{https://www.cosmos.esa.int/web/gaia/dpac/consortium}). Funding for the DPAC has been provided by national institutions, in particular the institutions participating in the Gaia Multilateral Agreement.
\end{acknowledgements}

\bibliography{aa}
\bibliographystyle{aa}

\begin{appendix}
\nolinenumbers

\section{Optical morphology and VLBI fields of view}\label{appendix_A}
\begin{figure}[htbp]
\centering
\includegraphics[width=0.45\textwidth]{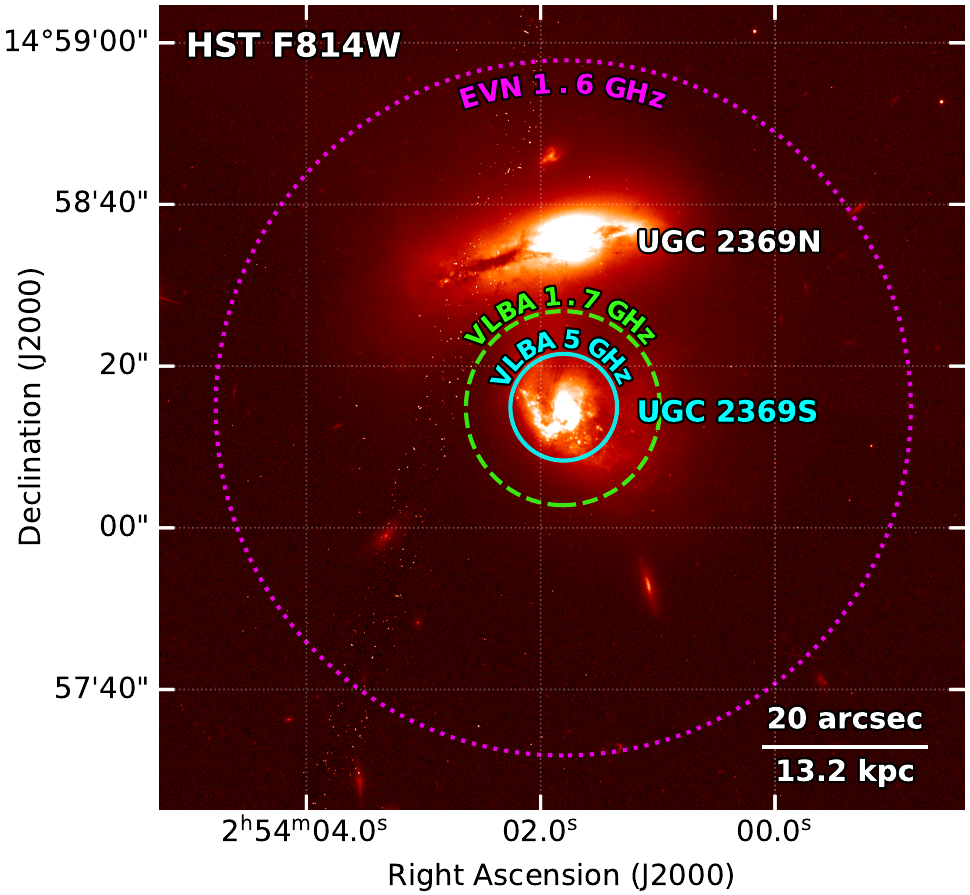}
\caption{HST/ACS F814W image of the interacting galaxy pair UGC~2369. The cyan, green, and magenta circles represent the effective FoVs for the VLBA 5~GHz, VLBA 1.7~GHz, and EVN 1.6~GHz (comprising eight European antennas) observations, respectively, with radii of $6.6\arcsec$, $12.0\arcsec$, and $43.0\arcsec$. These FoVs are centered on their actual phase centers, within which the amplitude loss due to smearing is $<10\%$ (estimated using the EVN Observation Planner, with 0.5~MHz channels and 2~s integration). The target highlighted in this work is the southern triple-merger luminous infrared galaxy UGC\,2369S \citep{2013ApJ...768..102K}.}
\label{figure_A1}
\end{figure}

Figure~\ref{figure_A1} shows the effective fields of view (FoVs) of the VLBI observations presented in this paper, overlaid on the Hubble Space Telescope (HST) F814W image of UGC~2369 (HST Proposal 10592, PI: A. S. Evans; \citealt{2013ApJ...768..102K}).

\section{Astrometric offsets of the radio and optical cores}\label{appendix_B}

Due to the update of the phase calibrator positions, we applied an astrometric correction to the model coordinates of the target source based on the latest data release (RFC\_2026a) of the Astrogeo Radio Fundamental Catalogue \citep{petrov2025}. The corrections applied to the phase calibrators are listed in Table~\ref{table_B1}. Consequently, the absolute coordinates of the target were shifted by $(\Delta \alpha \cos \delta,\Delta \delta) = (-0.70, -1.36)~{\rm mas}$ for the EVN 1.6~GHz observation, and $(+0.04,-4.46)~{\rm mas}$ and $(-0.62, -1.19)~{\rm mas}$ for the VLBA 1.7~GHz and 5~GHz observations, respectively.

\begin{table}[htbp]
\centering
\small
\setlength{\tabcolsep}{5.8pt}
\begin{threeparttable}
\caption{Astrometric corrections of the phase calibrators.} 
\label{table_B1}
\begin{tabular}{cccc}
\toprule\midrule
Project & Calibrator & Observation Pos. & Corrected Pos.  \\
        &            & (J2000)          & (J2000)         \\
\midrule
EC020B & J0256+1334 & \makecell{$02^{\rm h}56^{\rm m}34\fs9847$ \\[-1.5pt] +13\degr34\arcmin35\farcs346} & \makecell{$02^{\rm h}56^{\rm m}34\fs984652$ \\[-1.5pt] +13\degr34\arcmin35\farcs34464}  \\[6pt]
BL029  & J0238+1636  & \makecell{$02^{\rm h}38^{\rm m}38\fs9301$ \\[-1.5pt] +16\degr36\arcmin59\farcs297} & \makecell{$02^{\rm h}38^{\rm m}38\fs930103$ \\[-1.5pt] +16\degr36\arcmin59\farcs29454}  \\[6pt]
BM190A & J0256+1334 & \makecell{$02^{\rm h}56^{\rm m}34\fs984695$ \\[-1.5pt] +13\degr34\arcmin35\farcs34583} & \makecell{$02^{\rm h}56^{\rm m}34\fs984652$ \\[-1.5pt] +13\degr34\arcmin35\farcs34464}  \\  
\bottomrule
\end{tabular}
\begin{tablenotes}
\item {\bf Note.} The corrected coordinates and their uncertainties for the phase calibrators are taken from the Astrogeo RFC\_2026a catalogue \citep{petrov2025}. The positional uncertainties are $\pm 0.22$~mas for J0256+1334 and $\pm 0.15$~mas for 0235+164.
\end{tablenotes}
\end{threeparttable}
\end{table}

\begin{figure}[htbp]
\centering
\includegraphics[width=0.45\textwidth]{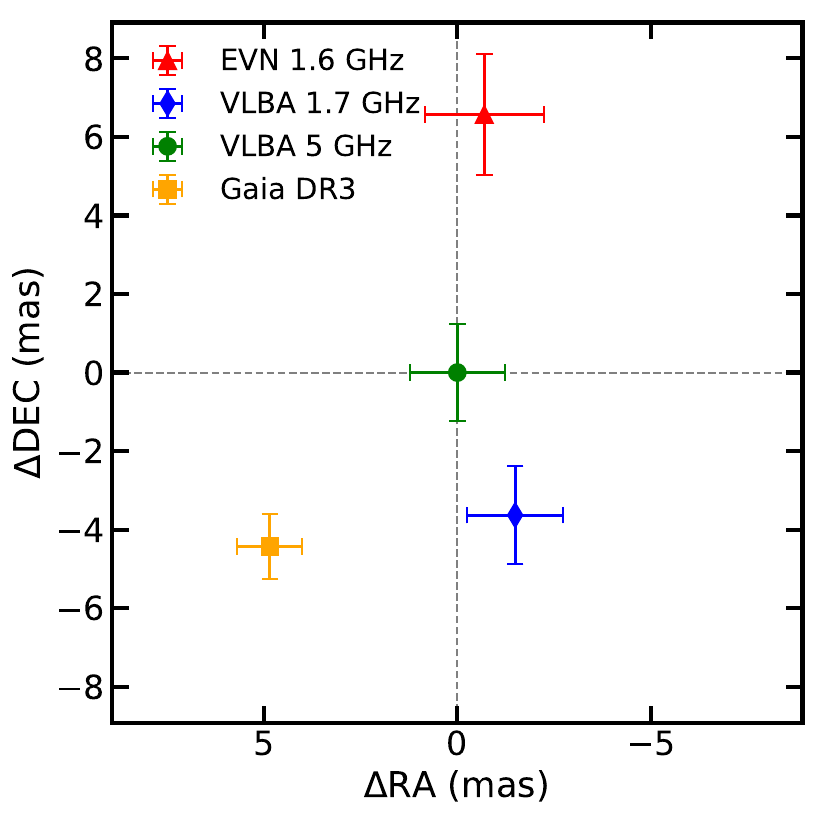}
\caption{Relative positional offsets of the northern core between the Gaia DR3 and multi-frequency VLBI observations. The phase-referenced position of the VLBA 5~GHz core is set as the origin (0,~0). Error bars correspond to the positional uncertainties detailed in Table~\ref{table_2}.}
\label{figure_B1}
\end{figure}

The phase-referenced radio coordinates of the northern core exhibit large offsets ($\sim$10~mas) among the three VLBI observations, significantly exceeding the formal astrometric uncertainties (Figure~\ref{figure_B1}). These internal radio offsets likely arise from spatial resolution mismatches, as well as frequency- and epoch-dependent structural changes. The VLBA 5~GHz image provides the highest spatial resolution, and thus most likely represents the true position of the AGN jet base. In addition, the VLBI coordinates also exhibit a significant offset ($>$5~mas) from the Gaia DR3 optical position. This discrepancy can be explained by Gaia photocenter jitter caused by the host galaxy's starlight and the merger-driven dusty environment, given the irregular morphology of UGC~2369S and its significant astrometric excess noise (${\rm AEN}>7\sigma$) in the Gaia DR3 catalogue \citep{Gaia2016,Gaia2023}.

\section{Calculations of model parameters}\label{appendix_C}

The theoretical resolution limit of VLBI imaging, $d_\text{lim}$, adopted as the upper limit for the unresolved core angular sizes at 1.6 and 1.7~GHz, is calculated as \citep{2005astro.ph..3225L}:
\begin{equation}
d_\text{lim} = \frac{\pi}{2} \left[\pi B_{\rm maj} B_{\rm min}\ln2 \, \ln\left(\frac{{\rm SNR}}{{\rm SNR}-1}\right) \right]^{1/2},
\label{eq1}
\end{equation}
where $B_{\rm maj}$ and $B_{\rm min}$ are the major and minor axes of the synthesized beam, and SNR is the signal-to-noise ratio. For the partially resolved VLBA 5~GHz image, the Gaussian size uncertainty is estimated as $\sigma_\theta = \theta / {\rm SNR}$.

The brightness temperature of the core is derived using \citep{Condon1982}:
\begin{equation}
T_{\rm b} = 1.22 \times 10^{12} (1+z) \frac{S_{\nu}}{\nu^2 \theta^2}~~\rm [K],
\label{eq2}
\end{equation}
where $z$ is the redshift, $S_{\nu}$ is the flux density in the unit of jansky, $\nu$ is the observing frequency in gigahertz, $\theta$ is the circular Gaussian model angular diameter (full width at half-maximum, FWHM) in mas. 

Rest-frame monochromatic radio luminosities are estimated using the following equation \citep{hogg2002k}:
\begin{equation}
L_{\nu} = 4\pi d_{\rm L}^2 \frac{S_{\nu}}{(1+z)^{1+\alpha}},
\label{eq3}
\end{equation}
where $d_{\rm L}$ is the luminosity distance and $\alpha$ is the spectral index. Given the low redshift of UGC~2369S, we approximated the monochromatic luminosity as $L_{\nu}\approx4\pi d_{\rm L}^2 S_{\nu}$. 

\section{Multi-wavelength analysis}\label{appendix_D}

\begin{table*}[htbp]
\centering
\begin{threeparttable}
\caption{Multiband emission properties of the three cores in UGC~2369S.} 
\label{table_D1}
\begin{tabular}{ccccccc}
\toprule\midrule
Component & $M_{\rm BH}$ & $\log L_{\rm [O\,III]}^{\rm int}$ & $S_{\rm 5GHz}$ & $\log \nu L_{\nu}$ & $\log L_{\rm X,FP}$ & $\log L_{\rm 2-8keV}$ \\
    & ($M_{\odot}$) & (erg~s$^{-1}$) & (mJy) & (W) & (erg~s$^{-1}$) & (erg~s$^{-1}$) \\
(1) & (2) & (3) & (4) & (5) & (6) & (7)\\
\midrule
N  & $8.67\pm0.02$ & $>41.14$ & $2.10\pm0.23$ & $31.42\pm0.05$ & $42.00\pm0.41$ & $40.12\pm0.08$  \\
SE & $8.18\pm0.02$ & $>40.65$ & $<0.53$       & $<30.83$       & $<41.58$       & $<39.60$        \\
SW & $6.96\pm0.05^{\dagger}$  & $>40.20$ & $<0.53$ & $<30.82$  & $<42.64$       & $<39.48$        \\
\bottomrule
\end{tabular}
\begin{tablenotes}
\item {\bf Note.} Column~(1): component name (N: northern, SE: southeast, SW: southwest). Column~(2): black hole mass estimated from the stellar velocity dispersion. $^{\dagger}$The $M_{\rm BH}$ of SW core may be highly uncertain due to tidal stripping \citep{2026ApJ..1002..130D}. Column~(3): extinction-corrected intrinsic [O~III] luminosity. Column~(4): VLBA 5~GHz flux density or $6\sigma$ upper limit. Column~(5): VLBA 5~GHz radio luminosity or upper limit. Column~(6): expected X-ray luminosity or upper limit derived from the FP relation (Equation~\ref{eq4}). Column~(7): Chandra-observed X-ray luminosity or upper limit \citep{2026ApJ..1002..130D}.
\end{tablenotes}
\end{threeparttable}
\end{table*}

\subsection{Fundamental plane relation}\label{appendix_D1}

The fundamental plane (FP) of black hole activity connects the radio luminosity, X-ray luminosity, and black hole mass across a wide range of accreting systems, indicating a scale-invariant nature of the disk--jet coupling \citep{2003MNRAS.345.1057M,2004A&A...414..895F}. To evaluate the intrinsic X-ray luminosity from the radio emission, we adopt the Bayesian regression derived for sub-Eddington objects by \citet{2012MNRAS.419..267P}:
\begin{equation}
\begin{split}
\log L_{\rm X,FP} =\ & (1.45 \pm 0.04) \log L_{\rm R} - (0.88 \pm 0.06) \log M_{\rm BH} \\
&- 6.07 \pm 1.10,
\end{split}
\label{eq4}
\end{equation}
where $L_{\rm X,FP}$ and $L_{\rm R}$ are the intrinsic X-ray and radio luminosities in units of $\mathrm{erg~s^{-1}}$, respectively, and $M_{\rm BH}$ is the black hole mass in units of solar masses ($M_{\odot}$). The core radio luminosity is derived from the VLBA 5~GHz observations, yielding $L_{\rm R}=\nu L_{\nu}=(2.61\pm0.29)\times10^{31}~{\rm W} \equiv (2.61\pm0.29)\times 10^{38}~{\rm erg~s^{-1}}$. The black hole mass used in the calculation is $M_{\rm BH}=(4.73\pm0.24)\times10^8~M_{\odot}$ \citep{2026ApJ..1002..130D}, which was estimated from the stellar velocity dispersion ($\sigma_*$) via the $M_{\rm BH}-\sigma_*$ relation \citep{2013ARA&A..51..511K}. The final predicted uncertainty was calculated by adding the propagated observational errors in quadrature with the intrinsic scatter of the FP relation ($\sigma_{\rm int} \sim 0.40$~dex; \citealt{2012MNRAS.419..267P}). To assess the degree of X-ray obscuration, the hydrogen column density ($N_{\rm H}$) was estimated using standard photoelectric absorption models via WebPIMMS\footnote{\url{https://heasarc.gsfc.nasa.gov/cgi-bin/Tools/w3pimms/w3pimms.pl}}. For this calculation, we assumed a primary power-law X-ray spectrum with a typical photon index of $\Gamma = 1.8$ \citep[e.g.,][]{2006A&A...451..457T, 2017ApJS..233...17R}.

\subsection{Eddington ratio}\label{appendix_D2}

The Eddington ratio, which characterizes the accretion efficiency of the supermassive black hole, can be calculated as \citep{2008ARA&A..46..475H}:
\begin{equation}
\lambda_{\rm Edd}=\frac{L_{\rm bol}}{L_{\rm Edd}}=\frac{16~L_{\rm X}}{1.26\times10^{38}~M_{\rm BH}/M_{\odot}},
\label{eq5}
\end{equation}
where $L_{\rm bol}$ represents the bolometric luminosity, which is estimated from the intrinsic X-ray luminosity $L_{\rm X}$ (in units of $\rm erg~s^{-1}$) derived from the VLBA 5~GHz luminosity via the FP relation, and $L_{\rm Edd}$ is the Eddington luminosity. Assuming a typical correction factor of $k_{\rm bol} \approx 16$ for low-luminosity AGNs \citep{2008ARA&A..46..475H}, the intrinsic bolometric luminosity of the northern core in UGC~2369S is estimated as $\log(L_{\rm bol}/{\rm erg~s^{-1}}) = 43.20 \pm 0.41$. With a black hole mass of $M_{\rm BH}=(4.73\pm0.24)\times10^8M_{\odot}$ \citep{2026ApJ..1002..130D}, the Eddington luminosity and Eddington ratio are derived as $\log (L_{\rm Edd}/{\rm erg~s^{-1}}) = 46.78\pm0.02$ and $\lambda_{\rm Edd} = \left(2.68^{+4.22}_{-1.64}\right) \times 10^{-4}$, respectively.

\subsection{Thickness parameter}\label{appendix_D3}

The thickness parameter, defined as the ratio of the observed hard X-ray luminosity to the extinction-corrected [O~III] luminosity ($T=L_{\rm X}/L_{\rm [O\,III]}^{\rm int}$), is widely adopted as an empirical diagnostic for extreme nuclear obscuration \citep{1999ApJS..121..473B,2006A&A...455..173P}. For the northern core in UGC~2369S, the observed ${\rm [O~III]}~\lambda 5008$ emission line flux is $(4.02\pm0.01)\times10^{-15}~{\rm erg~s}^{-1}~{\rm cm}^{-2}$ within a $1\farcs5$ aperture \citep{2026ApJ..1002..130D}. Given the heavy dust obscuration in the nuclear region ($A_{\rm V}>5$, $N_{\rm H}\gtrsim10^{25}$~cm$^{-2}$; \citealt{2026ApJ..1002..130D}), we applied an extinction correction of $A_{5007} = k(5007) \times E(B-V) \approx 3.48\times 0.82 \approx 2.85$~mag based on the Balmer decrement \citep{1989ApJ...345..245C}, which yields an intrinsic luminosity of $L_{\rm [O\,III]}^{\rm int}\approx1.38\times10^{41}~{\rm erg~s}^{-1}$. Combining this with the Chandra-observed X-ray luminosity yields a remarkably low thickness parameter of $T \sim 0.09$.

\begin{figure}[htbp]
\centering
\includegraphics[width=0.45\textwidth]{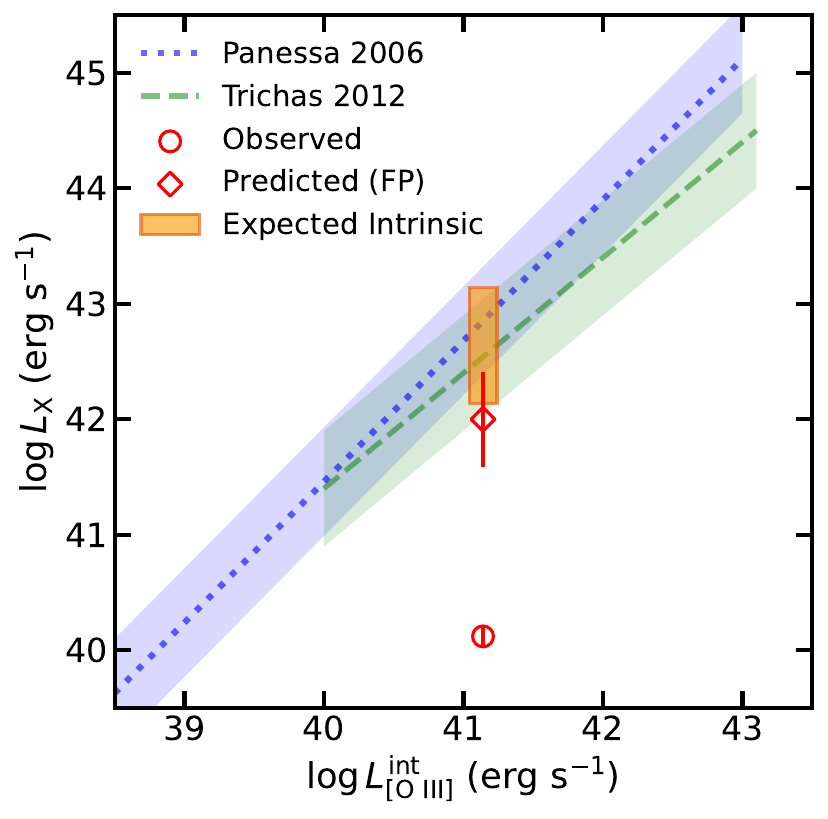}
\caption{X-ray luminosity vs. extinction-corrected intrinsic ${\rm [O~III]}~\lambda 5007$ luminosity. Purple dotted line: local Seyfert relation with a 0.47~dex intrinsic scatter \citep{2006A&A...455..173P}. Green dashed line: Type 2 Seyfert relation from the CSC-SDSS sample with a 0.5~dex intrinsic scatter \citep{2012ApJS..200...17T}.}
\label{figure_D1}
\end{figure}

As shown in Figure~\ref{figure_D1}, where the dotted and dashed lines represent the empirical relations for local \citep{2006A&A...455..173P} and Type 2 Seyferts \citep{2012ApJS..200...17T}, the observed X-ray emission (red circle) falls dramatically below these intrinsic expectations by $\gtrsim2$~dex. To quantitatively assess this, we derived an expected intrinsic X-ray luminosity from the extinction-corrected [O~III] luminosity, adopting a typical intrinsic $L_{\rm X}$ to $L_{\rm [O\,III]}^{\rm int}$ ratio of $10-100$ established for unobscured Type 1 AGNs \citep{2005ApJ...634..161H}:
\begin{equation}
\begin{split}
\log L_{\rm X,[O\,III]} =\log L_{\rm [O\,III]}^{\rm int} + 1.59 \pm 0.48.
\end{split}
\label{eq6}
\end{equation}
The consistency between the FP-predicted (red diamond) and [O~III]-derived (orange shaded region) intrinsic X-ray luminosities indicates a powerful central engine deeply hidden behind a Compton-thick envelope.

\subsection{Radio emission origins}\label{appendix_D4}

To investigate whether the radio emission of the northern core originates from the corona or a jet, we derive a radio-to-X-ray ratio of $R_{\rm X} = L_{\rm R} / L_{\rm X}^{\rm int}\approx 10^{-4.22}$, using the VLBA 5~GHz luminosity and intrinsic $L_{\rm X}$ estimated from the [O~III] emission. This value is nearly an order of magnitude higher than the G\"udel--Benz relation ($\sim 10^{-5}$; \citealt{1993ApJ...405L..63G,2008MNRAS.390..847L}), indicating that the radio emission detected by VLBI is jet-dominated. 

Adopting the star formation rate (SFR) of $14.70 \pm 1.53~M_{\odot}~\rm yr^{-1}$ for the northern core \citep{2026ApJ..1002..130D}, estimated from the VLA C-configuration 33~GHz flux density of $1.46\pm0.15$~mJy within a $1\farcs5$ aperture \citep{2019ApJ...881...70L,2022ApJ...940...52S} based on the empirical calibrations of \citet{2011ApJ...737...67M, 2012ApJ...761...97M} with an assumed spectral index of $\alpha=-0.85$ ($S_{\nu}\propto\nu^{\alpha}$), we calculate the expected 1.4~GHz luminosity contributed by star formation as follows:
\begin{equation}
\left(\frac{\rm SFR}{M_{\odot}\,{\rm yr^{-1}}}\right) =6.35\times10^{-29}\left(\frac{L_{\rm 1.4\,GHz}}{\rm erg\,s^{-1}\,Hz^{-1}}\right).
\label{eq7}
\end{equation}
This yields $L_{\rm 1.4\,GHz,SF}=(2.32\pm0.24)\times10^{22}~{\rm W~Hz^{-1}}$, corresponding to an expected 1.4~GHz flux density of $9.3 \pm 1.0~{\rm mJy}$. Considering the VLA A-configuration 1.49~GHz flux density of $41.9\pm0.2$~mJy within a $1\farcs5$ ($\sim1~{\rm kpc}$) region \citep{1990ApJS...73..359C,2015A&A...574A...4V}, the remaining $\sim 33~{\rm mJy}$ of radio emission cannot be attributed to an aperture mismatch, and instead likely originates from non-star-forming processes, such as AGN jets, outflows (winds), or coronal activity \citep{2019NatAs...3..387P}. Specifically, jet-dominated compact radio emission (including a minor contribution from the corona) with a flux density of $4.80\pm0.51~{\rm mJy}$ was detected in the EVN 1.6~GHz observations, concentrated within a 12~mas ($\sim 7.9~{\rm pc}$) core. The residual $\sim 28~{\rm mJy}$ of non-thermal emission, which is resolved out by VLBI but contained within the VLA beam, likely originates from extended jet components or AGN-driven outflows.

\end{appendix}

\end{document}